\documentstyle[aps,eqsecnum,amsfonts,preprint,psfig]{revtex}
\begin{document}
\draft
\title{\hfill OKHEP-93-05\\
Casimir effect for a $D$-dimensional sphere}
\author {Carl M. Bender\thanks{E-mail: cmb@howdy.wustl.edu}}
\address{Department of Physics,
	Washington University,
	St.\ Louis, MO 63130}
\author{Kimball A. Milton\thanks{E-mail:
milton@phyast.nhn.uoknor.edu}}
\address{Department of Physics and Astronomy,
	University of Oklahoma,
	Norman, OK 73019}
\date{\today}
\maketitle
\begin{abstract}
The Casimir force on a $D$-dimensional sphere due to the confinement
of a
massless scalar field is computed as a function of $D$, where $D$ is
a
continuous variable that ranges from $-\infty$ to $\infty$. The
dependence of
the force on the dimension is obtained using a simple and
straightforward
Green's function technique. We find that the Casimir force vanishes
as
$D\to +\infty$ ($D$ non-even integer) and also vanishes when $D$ is a
negative
even integer. The force has simple poles at positive even integer
values of $D$.
\end{abstract}

\pacs{11.10.Kk, 12.39.Ba, 11.10.Jj}

\section{Introduction}
\label{sec:intro}
In recent papers \cite{BBL1,BBL2,Fermitalk,finf} it was proposed that
the
dimension of space-time could be used as a perturbation parameter in
quantum
field theory calculations. The advantage of such an approach is that
analytical
(nonnumerical) results can be obtained which are nonperturbative in
the coupling
constant. This procedure was used to obtain the Green's functions of
self-interacting scalar quantum field theory in the Ising limit
\cite{Ising1,Ising2}. One can also perform dimensional expansions in
inverse
powers of the dimension. Such expansions have proved useful in atomic
physics
calculations \cite{dinv}.

These perturbative investigations have led to and motivated analyses
of the
dimensional dependence of various physical systems. Such
investigations are
useful because by identifying the singularities in the
complex-dimension plane
one can predict the radius of convergence of a dimensional expansion.
The
dimensional dependence of some elementary quantum-mechanical and
field-theoretic
models is described in \cite{BBL1,BBL2,Fermitalk}. The dimensional
dependence of
classical physical models has also been investigated; for example, in
\cite{walk1,walk2} the dimensional dependence of probabilities in
models of
random walks was elucidated.

In this paper we investigate the dimensional dependence of the
Casimir force per
unit area, $F/A$, on a spherical shell of radius $a$ in $D$ space
dimensions.
Specifically, we study the Casimir force that is due to quantum
fluctuations of
a free massless scalar field satisfying Dirichlet boundary conditions
on the
shell.

An interesting investigation of the dependence of the Casimir force
per unit
area upon the spatial dimension is already in the literature
\cite{wolf}.
Ambj{\o}rn and Wolfram examined the case of infinite parallel plates
embedded in
a $D$-dimensional space and separated by a distance $2a$; that is,
there is one
longitudinal dimension and $D-1$ transverse dimensions. Their result
is
\begin{equation}
F/A=-a^{-D-1}2^{-2D-2}\pi^{-(D+1)/2}D\,\Gamma\left ({D+1\over
2}\right )
\zeta (D+1),
\label{0}
\end{equation}
which we have plotted in Fig.~1. Note that $F/A$ has a simple pole
(due to the
Gamma function) at $D=-1$. However, $F/A$ is not infinite at the
other poles of
the Gamma function, which are located at all the negative odd
integral values of
$D$ because the Riemann zeta function vanishes at all negative even
values of
its argument. One interesting and well-known special case of
(\ref{0}) is $D=1$:
\begin{equation}
F/A|_{D=1} =-{\pi\over 96 a^2},
\label{00}
\end{equation}
where the negative sign indicates that the force is attractive.
We mention this case here because the spherical geometry considered
in the
present paper coincides with the slab geometry of \cite{wolf} when
$D=1$; we
recover the result (\ref{00}) as a special case in Sec.~\ref{numer}.

This paper is organized very simply. In Sec.~\ref{formalism} we
review the
Green's function formalism required to obtain the Casimir force. Then
we apply
this formalism in $D$-dimensional space to obtain an expression for
the Casimir
force per unit area on a $D$-dimensional spherical shell. This
expression takes
the form of an infinite sum of integrals over modified Bessel
functions; the
dimension $D$ appears in the orders of the Bessel functions. In
Sec.~\ref{numer}
we examine this expression for the Casimir force per unit area in
detail. We
show that each term in the series exists (each of the integrals
converges) and
we show how to evaluate the sum of the series numerically for all
real $D$. When
$D>0$ the Casimir force per unit area is real; the force is finite
except when
$D$ is an even integer. When $D\leq 0$ the Casimir force is complex;
there are
logarithmic singularities in the complex-$D$ plane at
$D=0,~-2,~-4,~-6,~\ldots$.

\section{Mathematical formalism}
\label{formalism}
The calculation in this paper of the Casimir force on a spherical
shell relies
on the use of Green's functions to represent vacuum expectation
values of
time-ordered products of fields. The Green's function is used to
obtain the
vacuum expectation value of the stress-energy tensor, from which we
will derive
the Casimir force. The formalism used here was developed in
\cite{mil1,mil2,mil3}. We summarize the formalism below.

A free massless scalar field $\varphi ({\bf x},t)$ satisfies the
Klein-Gordon
equation
\begin{equation}
\left ({{\partial^2}\over{\partial t^2}}-\nabla^2\right )\varphi({\bf
x},t)=0,
\label{1}
\end{equation}
where ${\bf x}$ is a $D$-dimensional position vector. (Initially, we
will think
of $D$ as a positive integer; however, once we have derived the
radial equation
for the Green's function we will be able to regard the parameter $D$
as a
continuous variable.) The quantum nature of the Casimir force arises
from the
constraint that $\varphi({\bf x},t)$ satisfies equal-time commutation
relations:
\begin{equation}
[\varphi({\bf x},t),\dot\varphi ({\bf x}',t)]=i\delta^{(D)}({\bf
x}-{\bf x}').
\label{2}
\end{equation}

The two-point Green's function $G ({\bf x},t; {\bf x}',t')$ is
defined as the
vacuum expectation value of the time-ordered product of two fields:
\begin{equation}
G({\bf x},t;{\bf x}',t')\equiv -i\langle 0 |{\rm T}\varphi({\bf x},t)
\varphi({\bf y},t') | 0 \rangle.
\label{3}
\end{equation}
By virtue of (\ref{1}) and (\ref{2}), the Green's function
$G({\bf x},t;{\bf x}',t')$ satisfies the inhomogeneous Klein-Gordon
equation
\begin{equation}
\left ({{\partial^2}\over{\partial t^2}}-\nabla^2\right ) G({\bf
x},t;{\bf x}',
t') =-\delta^{(D)} ({\bf x}-{\bf x}')\delta (t-t').
\label{4}
\end{equation}
We will solve the above Green's function equation by dividing space
into two
regions, {\sl region I}, the interior of a sphere of radius $a$ and
{\sl region
II}, the exterior of the sphere. On the sphere we will impose
Dirichlet boundary
conditions
\begin{equation}
G({\bf x},t;{\bf x}',t')\bigm | _{|{\bf x}|=a}=0.
\label{5}
\end{equation}
In addition, in {\sl region I\/} we will require that $G$ be finite
at the
origin ${\bf x}=0$ and in {\sl region II\/} we will require that $G$
satisfy
outgoing-wave boundary conditions at $|{\bf x}|=\infty$.

The stress-energy tensor $T^{\mu\nu}({\bf x},t)$ is defined as
\cite{elcid}
\begin{equation}
T^{\mu\nu}({\bf x},t)\equiv\partial^{\mu}\varphi({\bf
x},t)\partial^{\nu}\varphi
({\bf x},t)-{1\over 2}g^{\mu\nu}\partial_{\lambda}\varphi({\bf
x},t)\partial^
{\lambda}\varphi({\bf x},t).
\label{6}
\end{equation}
The radial Casimir force per unit area $F/A$ on the sphere is
obtained from the
radial-radial component of the vacuum expectation value of the
stress-energy
tensor \cite{mil1}:
\begin{equation}
F/A =\langle 0|T^{rr}_{\rm in}-T^{rr}_{\rm out}|0\rangle\bigm
|_{r=a}.
\label{7}
\end{equation}
To calculate $F/A$ we exploit the connection between the vacuum
expectation
value of the stress-energy tensor $T^{\mu\nu} ({\bf x},t)$ and the
Green's
function at equal times $G({\bf x},t;{\bf x}',t)$:
\begin{equation}
F/A ={i\over 2}\left [{\partial\over\partial r}{\partial\over\partial
r'}G({\bf
x},t;{\bf x}',t)_{\rm in}-{\partial\over\partial
r}{\partial\over\partial r'}
G({\bf x},t;{\bf x}',t)_{\rm out}\right]\Bigg |_{{\bf x}={\bf
x}',~|{\bf x}|=a}.
\label{8}
\end{equation}

To evaluate the expression in (\ref{8}) it is necessary to solve the
Green's
function equation (\ref{4}). We begin by taking the time Fourier
transform of
$G$:
\begin{equation}
{\cal G}_\omega ({\bf x};{\bf
x}')=\int_{-\infty}^{\infty}dt\,e^{-i\omega
(t-t')} G({\bf x},t;{\bf x}',t').
\label{9}
\end{equation}
The differential equation satisfied by ${\cal G}_\omega$ is
\begin{equation}
\left ( \omega^2+\nabla^2 \right ) {\cal G}_\omega ({\bf x};{\bf x}')
= \delta^{(D)} ({\bf x}-{\bf x}').
\label{10}
\end{equation}

To solve this equation we introduce polar coordinates and seek a
solution that
has cylindrical symmetry; i.e., we seek a solution that is a function
only of
the two variables $r=|{\bf x}|$ and $\theta$, the angle between ${\bf
x}$ and
${\bf x}'$ so that ${\bf x}\cdot{\bf x'}= r r'\cos\theta$. In terms
of these
polar variables (\ref{10}) becomes
\begin{equation}
\left (\omega^2+{\partial^2\over\partial r^2}+{D-1\over
r}{\partial\over\partial
r}+{\sin^{2-D}\theta\over
r^2}{\partial\over\partial\theta}\sin^{D-2}\theta
{\partial\over\partial\theta}\right ){\cal G}_\omega
(r,r',\theta)={\delta(r-r')
\delta(\theta)\Gamma\left ( {D-1\over 2}\right )\over 2\pi^{(D-1)/2}
r^{D-1}
\sin^{D-2} \theta }.
\label{11}
\end{equation}
Note that the $D$-dimensional delta function on the right side of
(\ref{10}) has
been replaced by a cylindrically-symmetric delta function having the
property
that its volume integral in $D$ dimensional space is unity. The
$D$-dimensional
volume integral of a cylindrically-symmetric function $f(r,\theta)$
is
\begin{equation}
{2\pi^{(D-1)/2}\over \Gamma\left ( {D-1\over 2}\right )}
\int_0^{\infty}dr\, r^{D-1}\int_0^\pi d\theta \,\sin^{D-2} \theta
f(r,\theta).
\label{12}
\end{equation}

We solve (\ref{11}) using the method of separation of variables. Let
\begin{equation}
{\cal G}_\omega (r,r', \theta)= A(r) B(z),
\label{13}
\end{equation}
where $z=\cos\theta$. The equation satisfied by $B(z)$ is then
\begin{equation}
\left [ (1-z^2){d^2\over dz^2}-z(D-1){d\over dz}+n(n+D-2)\right ]
B(z)=0,
\label{14}
\end{equation}
where we have anticipated a convenient form for the separation
constant. The
equation satisfied by $A(r)$ is
\begin{equation}
\left [ {d^2\over dr^2}-{D-1 \over r}{d\over dr}-{n(n+D-2)\over r^2}
+
\omega^2 \right ] A(r)=0 \quad (r\neq r').
\label{15}
\end{equation}
The solution to (\ref{14}) that is regular at $|z|=1$ is the
ultraspherical
(Gegenbauer) polynomial \cite{NBS}
\begin{equation}
B(z)= C_n ^{(-1+D/2)} (z) \quad (n=0,\,1,\,2,\,3,\,\ldots).
\label{16}
\end{equation}
The solution in {\sl region I\/} to (\ref{15}) that is regular at
$r=0$ involves
the Bessel function \cite{NBS2}
\begin{equation}
A(r)= r^{1-D/2} J_{n-1+{D\over 2}}(|\omega| r).
\label{17}
\end{equation}
In (\ref{17}) we assume that $D\geq 2$ in order to eliminate the
linearly
independent solution $A(r)=r^{1-D/2}Y_{n-1+{D\over 2}}(|\omega| r)$,
which is
singular at $r=0$ for all $n$. The solution in {\sl region II\/} to
(\ref{15})
that corresponds to an outgoing wave at $r=\infty$ involves a Hankel
function of
the first kind \cite{NBS2}
\begin{equation}
A(r)=r^{1-D/2} H_{n-1+{D\over 2}}^{(1)} (|\omega| r).
\label{18}
\end{equation}

The general solution to (\ref{11}) is an arbitrary linear combination
of
separated-variable solutions; in {\sl region I\/} the Green's
function has the
form
\begin{mathletters}
\begin{equation}
{\cal G}_\omega (r,r', \theta)= \sum_{n=0}^\infty
a_n r^{1-D/2} J_{n-1+{D\over 2}}(|\omega| r) C_n ^{(-1+D/2)}(z)\quad
(r<r'<a)
\label{19a}
\end{equation}
and
\begin{equation}
{\cal G}_\omega (r,r',\theta)=\sum_{n=0}^\infty r^{1-D/2}\left [ b_n
J_{n-1+
{D\over 2}}(|\omega| r)+c_n J_{-n+1-{D\over 2}}(|\omega| r)
\right ] C_n ^{(-1+D/2)}(z)\quad (r'<r<a).
\label{19b}
\end{equation}
\end{mathletters}
[Note that $J_\nu (x)$ and $J_{-\nu} (x)$ are linearly independent so
long as
$\nu$ is not an integer. Thus, (\ref{19b}) assumes explicitly that
$D$ is not an
even integer.] The general solution to (\ref{11}) in {\sl region
II\/} has the
form
\begin{mathletters}
\begin{equation}
{\cal G}_\omega (r,r',\theta)=\sum_{n=0}^\infty d_n
r^{1-D/2}H^{(1)}_{n-1+{D
\over 2}}(|\omega| r) C_n ^{(-1+D/2)}(z)\quad (r>r'>a)
\label{20a}
\end{equation}
and
\begin{equation}
{\cal G}_\omega (r,r',\theta)=\sum_{n=0}^\infty r^{1-D/2}
\left [  e_n H^{(1)}_{n-1+{D\over 2} }(|\omega| r)+f_n
H^{(2)}_{n-1+{D\over 2}}
(|\omega| r)\right ] C_n ^{(-1+{D/2})}(z)\quad (r'>r>a).
\label{20b}
\end{equation}
\end{mathletters}

The arbitrary coefficients $a_n$, $b_n$, $c_n$, $d_n$, $e_n$, and
$f_n$ are
uniquely determined by six conditions; namely, the Dirichlet boundary
condition
(\ref{5}) at $r=a$,
\begin{mathletters}
\begin{equation}
b_n J_{n-1+{D\over 2}}(|\omega| a) + c_n J_{-n+1-{D\over 2}}(|\omega|
a) = 0
\label{21a}
\end{equation}
and
\begin{equation}
e_n H^{(1)}_{n-1+{D\over 2}}(|\omega| a)+f_n H^{(2)}_{n-1+{D\over
2}}(|
\omega| a)=0,
\label{21b}
\end{equation}
the condition of continuity at $r=r'$,
\begin{equation}
a_n J_{n-1+{D\over 2}}(|\omega| r') =
b_n J_{n-1+{D\over 2}}(|\omega| r') + c_n J_{-n+1-{D\over
2}}(|\omega| r')
\label{21c}
\end{equation}
and
\begin{equation}
d_n H^{(1)}_{n-1+{D\over 2}}(|\omega| r')=e_n H^{(1)}_{n-1+{D\over
2}}(|\omega|
r')+f_n H^{(2)}_{n-1+{D\over 2}}(|\omega| r'),
\label{21d}
\end{equation}
and the jump condition in the first derivative of the Green's
function at
$r=r'$,
\begin{equation}
b_n J'_{n-1+{D\over 2}}(|\omega| r')+c_n J'_{-n+1-{D\over
2}}(|\omega| r')-a_n
J'_{n-1+{D\over 2}}(|\omega| r')=
{(2n+D-2)\Gamma\left ({D-2\over 2}\right )\over 4 (\pi r')^{D\over 2}
|\omega|}
\label{21e}
\end{equation}
and
\begin{equation}
e_n H^{(1)\prime}_{n-1+{D\over 2}}(|\omega| r')+f_n
H^{(2)\prime}_{n-1+{D
\over 2}}(|\omega| r')-d_n H^{(1)\prime}_{n-1+{D\over 2}}(|\omega|
r')=
{(2n+D-2)\Gamma\left ({D-2\over 2}\right )\over 4 (\pi r')^{D\over 2}
|\omega|}.
\label{21f}
\end{equation}
\end{mathletters}
Here we have used the orthogonality property of the ultraspherical
polynomials
\cite{NBS}
\begin{equation}
\int_{-1}^1 dz\, (1-z^2)^{\alpha -1/2} C_n^{(\alpha)}(z)
C_m^{(\alpha)}(z)
={ 2^{1-2\alpha} \pi \Gamma(n+2\alpha) \over
n!\, (n+\alpha) \Gamma^2(\alpha)}\delta_{nm} \quad (\alpha\neq 0) ,
\label{22}
\end{equation}
the value of the ultraspherical polynomials at $z=1$,
\begin{equation}
C_n^{(\alpha)}(1)={\Gamma(n+2\alpha)\over n!\,\Gamma(2\alpha)}
\quad (\alpha\neq 0),
\label{23}
\end{equation}
and the duplication formula,
$\Gamma(2x)=2^{2x-1}\Gamma(x)\Gamma(x+1/2)/
\sqrt{\pi}$.

Having determined the coefficients in the expressions for the Green's
function,
we can immediately evaluate the right side of (\ref{8}). The
contribution to
$F/A$ from the interior region ({\sl region I}) is
\begin{mathletters}
\label{24}
\begin{equation}
(F/A)_{\rm in}=i\sum_{n=0}^{\infty}{(n-1+{D\over 2})\Gamma
(n+D-2)\over 2^D
\pi^{D+1\over 2}a^D n!\,\Gamma\left ({D-1\over 2}\right
)}\int_{-\infty}^{
\infty}d\omega\left [ {|\omega|a J'_{n-1+{D\over 2}}(|\omega| a)\over
J_{n-1+{D\over 2}}(|\omega| a)}+1-{D\over 2} \right ].
\label{24a}
\end{equation}
The contribution to $F/A$ from the exterior region ({\sl region II})
is
\begin{equation}
(F/A)_{\rm out}=i\sum_{n=0}^{\infty}{(n-1+{D\over
2})\Gamma(n+D-2)\over 2^D
\pi^{D+1\over 2}a^D n!\,\Gamma\left ({D-1\over 2}\right
)}\int_{-\infty}^{
\infty}d\omega\left [{|\omega|a H^{(1)\prime}_{n-1+{D\over
2}}(|\omega| a)\over
H^{(1)}_{n-1+{D\over 2}} (|\omega| a)} +1-{D\over 2} \right ].
\label{24b}
\end{equation}
\end{mathletters}

The integrals in (\ref{24}) are oscillatory and therefore very
difficult to
evaluate numerically. Thus, it is advantageous to perform a rotation
of 90
degrees in the complex-$\omega$ plane. The resulting final expression
for $F/A$
is
\begin{equation}
F/A=-\sum_{n=0}^{\infty}{(n-1+{D\over 2})\Gamma(n+D-2)\over
2^{D-1}\pi^{D+1\over
2}a^{D+1} n!\,\Gamma\left ({D-1\over 2}\right
)}\int_0^{\infty}dx\left [{x I'_
{n-1+{D\over 2}}(x)\over I_{n-1+{D\over 2}}(x)}+
{x K'_{n-1+{D\over 2}}(x)\over K_{n-1+{D\over 2}}(x)}+2-D \right ].
\label{25}
\end{equation}
The $D=2$ result, where the $n=0$ term appears with weight ${1\over
2}$, was
derived in \cite{mil3}. This result can be recovered by setting
$D=2+\epsilon$
and letting $\epsilon$ tend to $0$.

\section{Numerical evaluation of $F/A$}
\label{numer}
Our objective now is to evaluate the formal expression in (\ref{25})
for
arbitrary dimension $D$. Recall that this expression was derived
under the
assumption that $D>2$ and that $D$ is not an even integer. However,
we will now
seek an interpretation of (\ref{25}) that is generally valid; to do
so we will
apply a summation procedure that enables us to continue (\ref{25}) to
{\sl all\/} values of $D$.

The expression in (\ref{25}) does not exist {\sl a priori\/} for all
$D$.
Furthermore, as we will see, the individual terms in the series,
which are
integrals in $x$, do not exist. Fortunately, it is possible to modify
the
terms in the series so that the integrals do exist; this modification
requires a
delicate and detailed argument. However, there is one simple case,
namely that
for which $D=1$, where the series (\ref{25}) is well-defined and easy
to
evaluate. We examine this case in the next subsection.

\subsection{Special case $D=1$}
\label{sub1}
When $D=1$ the series in (\ref{25}) truncates after two terms. This
happens
because of the identity
\begin{equation}
\lim_{D\to 1} {\Gamma (n+D-2)\over \Gamma\left ({D-1\over 2}\right )}
=
-{1\over 2} \delta_{n0} + {1\over 2}\delta_{n1}.
\label{26}
\end{equation}
When this identity is inserted into the sum in (\ref{25}) we obtain
\begin{equation}
F/A|_{D=1}=-{1\over 4\pi a^2} \int_0^{\infty}dx \left [{x I'_{1\over
2}(x) \over
I_{1\over 2}(x)}+{x I'_{-{1\over 2}}(x)\over I_{-{1\over 2}}(x)}+{x
K'_{1\over
2}(x)\over K_{1\over 2}(x)}+{x K'_{-{1\over 2}}(x)\over K_{-{1\over
2}}(x)}+2
\right ].
\label{27}
\end{equation}
Next we use the identities
\begin{equation}
I_{1\over 2}(x) = {\sinh x\over \sqrt{x}}, \quad
I_{-{1\over 2}}(x) = {\cosh x\over \sqrt{x}},\quad
K_{1\over 2}(x) = K_{-{1\over 2}}(x) = {e^{-x}\over \sqrt{x}},
\label{28}
\end{equation}
to reduce (\ref{27}) to the (convergent) integral
\begin{eqnarray}
F/A|_{D=1} &=& -{1\over 4\pi a^2} \int_0^{\infty}dx \left [
x {d\over dx}\ln \left ( {\cosh x\sinh x\over x^2 e^{2x}}\right )
+2\right ]
\nonumber\\
&=& -{1\over 4\pi a^2}\int_0^{\infty}dx\, x{d\over dx}\ln\left(
1-e^{-4x}\right)
\nonumber\\
&=& -{1\over \pi a^2} \int_0^{\infty}dx {x\over e^{4x} -1 }
\nonumber\\
&=& -{\pi\over 96 a^2}.
\label{29}
\end{eqnarray}
This result agrees with the well-known result given in (\ref{00}).

\subsection{Convergent reformulation of (\protect\ref{25})}
\label{sub2}

In this subsection we modify the form of the series (\ref{25}) so
that each
term in the series exists and we apply a summation procedure to
evaluate the
the resulting series numerically. We begin by rewriting (\ref{25}) in
the form
\begin{equation}
F/A=-\sum_{n=0}^{\infty}{(n-1+{D\over 2})\Gamma (n+D-2)\over
2^{D-1}\pi^{D+1
\over 2}a^{D+1} n!\,\Gamma\left ({D-1\over 2}\right
)}\int_0^{\infty}dx\left [ x
{d\over dx}\ln\left ( I_{n-1+{D\over 2}}(x)K_{n-1+{D\over
2}}(x)\right ) +2-D
\right ].
\label{e4}
\end{equation}
In this form it is easy to investigate the convergence of the
individual
integrals in the series. To do so we recall the asymptotic behavior
as
$x\to\infty$:
\begin{equation}
I_\nu (x) K_\nu (x) \sim {1\over 2x}\quad (x\to +\infty).
\label{e5}
\end{equation}
 From (\ref{e5}) it is clear that the integrals in (\ref{e4}) do not
converge
except for the special case $D=1$. However, as we will now argue, one
can
replace the quantity $2-D$ in (\ref{e4}) by $1$ without changing the
value of
$F/A$, provided that $D<1$. This replacement will render the
integrals
convergent.

Consider the series
\begin{equation}
\sum_{n=0}^{\infty} {\Gamma(n+\alpha)\over n!}.
\label{e1}
\end{equation}
This series converges so long as $\alpha<0$ and $\alpha\neq -N$,
where
$N=1,~2,~3,~\ldots$ (so that individual terms in the series exist).
This series
can be summed in closed form because it is a special case of the
binomial
expansion
\begin{equation}
\sum_{n=0}^{\infty}x^n{\Gamma(n+\alpha)\over
n!}=\Gamma(\alpha)(1-x)^{-\alpha}.
\label{e2}
\end{equation}
Note that if we let $x\to 1$ we obtain the identity
\begin{equation}
\sum_{n=0}^{\infty}{\Gamma(n+\alpha)\over n!}\equiv 0.
\label{f1}
\end{equation}

One can also show that the identity (\ref{f1}) holds in the limit as
$\alpha$
approaches a negative integer $-N$. To do so we let
$\alpha=-N+\epsilon$, where
$N=1,~2,~3,~\ldots$. We then decompose the series into two parts, the
first
whose terms are finite and the second whose terms diverge as
$\epsilon\to 0$:
\begin{equation}
\sum_{n=0}^{\infty}{n!\over (n+N+1)!}+{1\over\epsilon}\sum_{n=0}^N
{(-1)^n\over
(N-n)!\, \Gamma(1+n-\epsilon)}.
\label{f2}
\end{equation}
The first sum in (\ref{f2}) can be easily evaluated as
\begin{equation}
{1\over N\, N!}.
\label{f3}
\end{equation}
If the second sum in (\ref{f2}) is expanded in a Laurent series in
$\epsilon$,
the coefficient of $1/\epsilon$ vanishes. However, the coefficient of
$\epsilon^0$ is given by
\begin{equation}
{1\over N!}\sum_{n=0}^N (-1)^n \left ( N \atop n \right ) \psi(n+1).
\label{f4}
\end{equation}
Finally, using the integral representation
\begin{equation}
\psi(n+1) = -\gamma + \int_0^1 dt {1-t^n\over 1-t},
\label{f5}
\end{equation}
we show that (\ref{f4}) cancels (\ref{f3}).

This argument shows that if the integral in (\ref{e4}) is replaced by
$1$
then the sum vanishes:
\begin{equation}
\sum_{n=0}^{\infty} {(n-1+{D\over 2}) \Gamma(n+D-2)\over n!}\equiv
0\quad (D<1).
\label{e3}
\end{equation}
It follows that in the region $D<1$ (if we sum first over $n$) we can
add any
constant to the integrand in each term in the series (\ref{e4})
without changing
the value of the sum. We conclude that we may replace $2-D$ by $1$ in
(\ref{e4}). Our new improved expression for the Casimir force per
unit area is
thus
\begin{equation}
F/A=\sum_{n=0}^{\infty}{(n-1+{D\over 2})\Gamma (n+D-2)\over
2^{D-1}\pi^{D+1
\over 2}a^{D+1} n!\,\Gamma\left ({D-1\over 2}\right
)}\int_0^{\infty}dx\,
\ln\left ( 2x I_{n-1+{D\over 2}}(x)K_{n-1+{D\over 2}}(x)\right ).
\label{f6}
\end{equation}
 For $D\geq 0$ each term in this series exists for all $n$. (For
$D<0$ there is
yet another subtlety that we will address shortly.) Before we
proceed, we must
emphasize that while (\ref{f6}) has a different and more compact form
than that
in (\ref{25}) we have not changed the value of $F/A$; we have in
effect added
zero to the series representing $F/A$.

Unfortunately, the formula in (\ref{f6}) is still not satisfactory
because the
series does not converge. To examine the convergence of this series
we need to
know the asymptotic behavior of the integrals for large $n$. We make
use of the
uniform asymptotic approximation to the product $I_\nu (\nu x)K_\nu
(\nu x)$:
\begin{equation}
I_\nu (\nu x)K_\nu (\nu x)\sim {t\over 2\nu}\left(
1+{t^2-6t^4+5t^6\over 8\nu^2}
+\ldots\right ) \quad (\nu\to\infty),
\label{e6}
\end{equation}
where $t=(1+x^2)^{-1/2}$. This asymptotic behavior implies that the
integral,
which we will abbreviate by $Q_n$, in the $n$th term in (\ref{f6})
grows
linearly with increasing $n$:
\begin{equation}
Q_n\equiv -\int_0^{\infty}dx\,\ln
[2xI_{\nu}(x)K_{\nu}(x)]\sim\pi\left ({\nu
\over 2}+{1\over 128\nu}-{35\over 32768\nu^3}+\ldots\right )\quad
(n\to\infty),
\label{f7}
\end{equation}
where $\nu=n-1+{D\over 2}$. Because of this linear growth in $n$ it
is apparent
that the series in (\ref{f6}) {\sl does not converge\/} if $D>0$
except for the
special case $D=1$, where the series truncates!

To solve this problem we introduce an analytic summation procedure
based on the
properties of the Riemann zeta function. Specifically, we consider
the leading
large-$n$ behavior of the summand in (\ref{f6}):
\begin{equation}
-{(n-1+{D\over 2})\Gamma (n+D-2)\over 2^{D-1}\pi^{D+1\over 2}a^{D+1}
n!\,\Gamma
\left ({D-1\over 2}\right )}Q_n \sim -{1\over 2^D \pi^{D-1 \over
2}a^{D+1}\Gamma
\left ({D-1\over 2}\right )}(n^{D-1} + \ldots ) \quad (n\to\infty).
\label{f8}
\end{equation}
Then, regarding the parameter $D$ as being less than $0$, we sum the
expression
on the right side of (\ref{f8}) over $n$ from $1$ to $\infty$. This
gives
\begin{equation}
-{1\over 2^D\pi^{D-1\over 2}a^{D+1}\Gamma\left ({D-1\over 2}\right
)}\zeta(1-D),
\label{f9}
\end{equation}
which is a well-defined function of $D$. We now add (\ref{f9}) to
(\ref{f6}) and
correspondingly subtract the right side of (\ref{f8}) from each term
(except the
$n=0$ term) in the expression for $F/A$ in (\ref{f6}). This produces
a new
series for $F/A$ that is convergent for $D<1$:
\begin{eqnarray}
F/A &=& {1\over a^{D+1} \pi^{D-1\over 2}}\Bigg \{
{\Gamma ({D\over 2})\over 4\pi^{3\over 2}}\int_0^{\infty}dx\,
\ln\left (2 x I_{-1+{D\over 2}}(x)K_{-1+{D\over 2}}(x)\right )
\nonumber\\
&& \quad + {1\over 2^{D-1}\pi\Gamma\left ({D-1\over 2} \right )}
\sum_{n=1}^{\infty} \Bigg [ {\pi\over 2}n^{D-1}
+{(n-1+{D\over 2})\Gamma (n+D-2) \over n!} \nonumber\\
&& \quad\quad\times\int_0^{\infty}dx\,
\ln\left ( 2x I_{n-1+{D\over 2}}(x)K_{n-1+{D\over 2}}(x)\right )
\Bigg ] -{1\over 2^D\Gamma\left ({D-1\over 2}\right )}\zeta(1-D)\Bigg
\}.
\label{f10}
\end{eqnarray}
We have finally achieved our objective; we have obtained a convergent
series
representation for $F/A$ for a finite range of positive $D$, namely,
$0<D<1$.
Each term in this series exists. In the next subsection we will use
this series
to calculate analytically the Casimir force on a sphere in zero
dimensions.

\subsection{Casimir force on a zero-dimensional sphere}
\label{sectionc}
If we substitute $D=0$ in (\ref{f10}) we find that $F/A=\infty$; this
is because
the Riemann zeta function $\zeta(1-D)$ is singular when $D=0$. The
divergence in
$F/A$ is a consequence of that fact that the surface area of a
zero-dimensional
sphere is zero. Hence, we will compute the Casimir force, rather than
the
Casimir force per unit area. The equation for $F$ is obtained by
multiplying
(\ref{f10}) by the surface area of a $D$-dimensional sphere of radius
$a$,
$A=2a^{D-1}\pi^{D/2}/\Gamma(D/2)$,
\begin{eqnarray}
F &=& { 1\over 2 \pi a^2 }\Bigg \{ \int_0^{\infty}dx\,
\ln\left ( 2x I_{-1+{D\over 2}}(x)K_{-1+{D\over 2}}(x)\right )
\nonumber\\
&& \quad + {2\over \Gamma (D-1)}\sum_{n=1}^{\infty} \Bigg [ {\pi\over
2}n^{D-1}
+{(n-1+{D\over 2})\Gamma (n+D-2) \over n!} \nonumber\\
&& \quad\quad\times\int_0^{\infty}dx\,
\ln\left ( 2x I_{n-1+{D\over 2}}(x)K_{n-1+{D\over 2}}(x)\right )
\Bigg ]
- {\pi\over\Gamma (D-1)}\zeta(1-D)\Bigg \}.
\label{h1}
\end{eqnarray}
We can now let $D$ tend to $0$ in (\ref{h1}). We obtain the result
\begin{equation}
F|_{D=0} = - {1\over 2 a^2},
\label{h2}
\end{equation}
where we have used $\zeta (z)\sim {1\over z-1}$ as $z\to 1$.

Note that we could not have obtained this result from (\ref{f6}).
Indeed, if
we naively let $D\to 0+$ in the formula obtained by multiplying
(\ref{f6}) by
the surface area $A$ of the sphere we appear to get the value $0$.
This is
because only the $n=0$ and $n=2$ terms survive in this limit and
these two terms
cancel as a result of the identities $I_{-n}(z)=I_n(z)$ and
$K_{-\nu}(z)=K_\nu
(z)$. However, the result $F=0$ at $D=0$ is incorrect because the
series in
(\ref{f6}) does not converge.

\subsection{Numerical results}
\label{sectiond}

The expression in (\ref{f10}) may in principle be used to compute
$F/A$
numerically for $D<1$; to wit, we may evaluate the integrals for a
large number
$N$ of terms in the series, compute the $N$th partial sum, and
extrapolate the
result to its value at $N=\infty$. However, this procedure is rather
inefficient
because the sum in (\ref{f10}) is very slowly converging. Therefore,
to prepare
for evaluating $F/A$ we subtract not just the one term in (\ref{f8})
but
many terms in this asymptotic expansion. The first three terms are
\begin{eqnarray}
&& -{(n-1+{D\over 2})\Gamma (n+D-2)\over 2^{D-1}\pi^{D+1\over
2}a^{D+1} n!\,
\Gamma \left ({D-1\over 2}\right )}Q_n \sim -{1\over 2^D
\pi^{D-1\over 2}a^{D+1}
\Gamma\left ({D-1\over 2}\right)}\bigg [ n^{D-1} +{(D-1)(D-2)\over
2}n^{D-2}
\nonumber\\
&& \quad\quad\quad +{24 D^4 - 176 D^3
+504 D^2 -688 D +387\over 192} n^{D-3} + \ldots \bigg ] \quad
(n\to\infty).
\label{f11}
\end{eqnarray}
If we use $K$ terms in this asymptotic expansion we then have $K$
corresponding
Riemann zeta functions appearing in the final form of the series. The
series
converges more rapidly (the $n$th term in the series vanishes like
$n^{D-K-1}$)
and it also converges for a larger range of the dimension: $D<K$. We
have used
this method to graph $F/A$ and $F$ as functions of $D$ (see Figs.~2
and 3).

 From Figs.~2 and 3 it appears that $F$ and $F/A$ are singular at
$D=2$ and
$D=4$. If fact, as we will now explain, the Casimir force is singular
at
{\sl all\/} even positive integer values of $D$; $F$ and $F/A$ have
simple poles
at $D=2N$, $N=1,~2,~3,~\ldots$. To verify this, we examine the
generalization
of (\ref{f10}) obtained by making many subtractions of the asymptotic
behavior
in (\ref{f11}). This formula for $F/A$ will contain many Riemann zeta
functions,
one for each subtraction. The $k$th zeta function will have the form
$\zeta(k-D)$. Furthermore, if $k$ is even the coefficient of
$\zeta(k-D)$
contains the factor $(k-1-D)$. (No such factor occurs if $k$ is odd.)
Thus, when
$D=k-1$ and $k$ is an even positive integer the simple pole of the
zeta function
is cancelled by this factor and when $D=k-1$ and $k$ is an odd
positive integer,
the simple pole persists.

As explained above, the Casimir force is finite at all odd-integer
dimensions.
For example,
\begin{equation}
F|_{D=3} = {1\over a^2} 0.0028168\ldots ,
\label{j1}
\end{equation}
where the positive value indicates that the force is repulsive (tends
to inflate
the sphere). The numerical value in (\ref{j1}) is much smaller than
that
obtained by Boyer \cite{tim} for the case of an electromagnetic field
confined
in a three-dimensional spherical cavity ($F=0.046176\ldots\, a^{-2}$)
and that
obtained by Milton \cite{kim} for the case of a spinor field confined
in a
three-dimensional spherical cavity ($F=0.0204\ldots\, a^{-2}$).

Although we have not proved it, it does appear from Fig.~3 that $F\to
0$ as
$D\to\infty$. This is probably associated with the fact that the
volume and
surface area of a $D$-dimensional sphere of radius $a$ tend to $0$ as
$D$
tends to $\infty$.

\subsection{Casimir force for negative dimension}
\label{sectione}

For all odd-integer $D\leq 1$ the series in (\ref{f6}) truncates and
thus it
is not necessary to subtract off the large-$n$ behavior. This
truncation
occurs because of the identity
\begin{equation}
\lim_{D\to 1-2N} {\Gamma (n+D-2)\over \Gamma\left ({D-1\over 2}\right
)} =
\sum_{j=0}^{2N+1} (-1)^{N+1+j}{N! \over 2 (2N+1-j)!}\delta_{nj} \quad
(N=0,\,1,\,2,\,3,\,\ldots),
\label{x26}
\end{equation}
which is the generalization of (\ref{26}). If we use this identity at
$N=1$ we
obtain the following integral representation for $F/A$ at $D=-1$:
\begin{equation}
F/A|_{D=-1} = -{1\over 2}\int_0^\infty dx\, \ln \left [
\left ( 1 + {1\over x} \right )^2 \left ( 1 - {\sinh x \over x\cosh
x}
\right ) \left ( 1 - {\cosh x \over x\sinh x } \right ) \right ],
\label{x514}
\end{equation}
where we have inserted the expressions for the half-odd integer
modified Bessel
functions. An interesting aspect of this integral representation for
$F/A$ is
that the argument of the logarithm has a zero for a positive real
value of $x$.
This zero may be traced to the positive zero of the function
$I_{-3/2}(x)$.
Thus, $F/A|_{D=-1}$ is complex!  Because the contour of integration
passes
under the zero, we find here $F/A|_{D=-1}=0.65382+i1.88445$.

In general, for all $D<0$, $D\neq -2N$ with $N=1,\,2,\,3,\,\ldots$,
the argument
of the logarithm in the integrand of (\ref{f6}) always has a zero.
Hence, the
analytic continuation of $F/A$ to negative values of $D$ is complex.
The zero of
the argument of the logarithm comes about because $I_{\nu}(x)$ has a
real
positive zero when $-2m<\nu<-2m+1$, where $m$ is a positive integer.
Only a
finite number of integrals in the series in (\ref{f6}) are complex.
However, as
$D$ becomes more negative, there are more and more complex integrals
in the
series. In particular, each time $D$ decreases past a negative even
integer
one additional integral in the series (\ref{f6}) becomes complex.
Thus, in
the complex-$D$ plane, $F/A$ has branch cuts emanating from the
points $D=-2k$,
$k=0,\,1,\,2,\,3,\,\ldots$.

It is remarkable, however, that exactly at the negative even
integers, it is
possible to evaluate the Casimir force $F$; we find that at these
points $F=0$.
This is because the series (\ref{f6}) truncates after a finite number
of terms
for these values of $D$, and the remaining terms cancel in pairs. In
Fig.~4
we plot ${\rm Re}\, F$ for $-5<D<5$. Figure 4 illustrates one
interesting aspect
of the Casimir force, namely, the erratic fluctuations in the sign of
$F$.
The sign of the Casimir force is extremely difficult to understand
intuitively---we know of no simple physical argument that predicts
whether the force is attractive or repulsive.
\bigskip
\bigskip

We thank the US Department of Energy for financial support.

\begin{figure}
\centerline{\psfig{figure=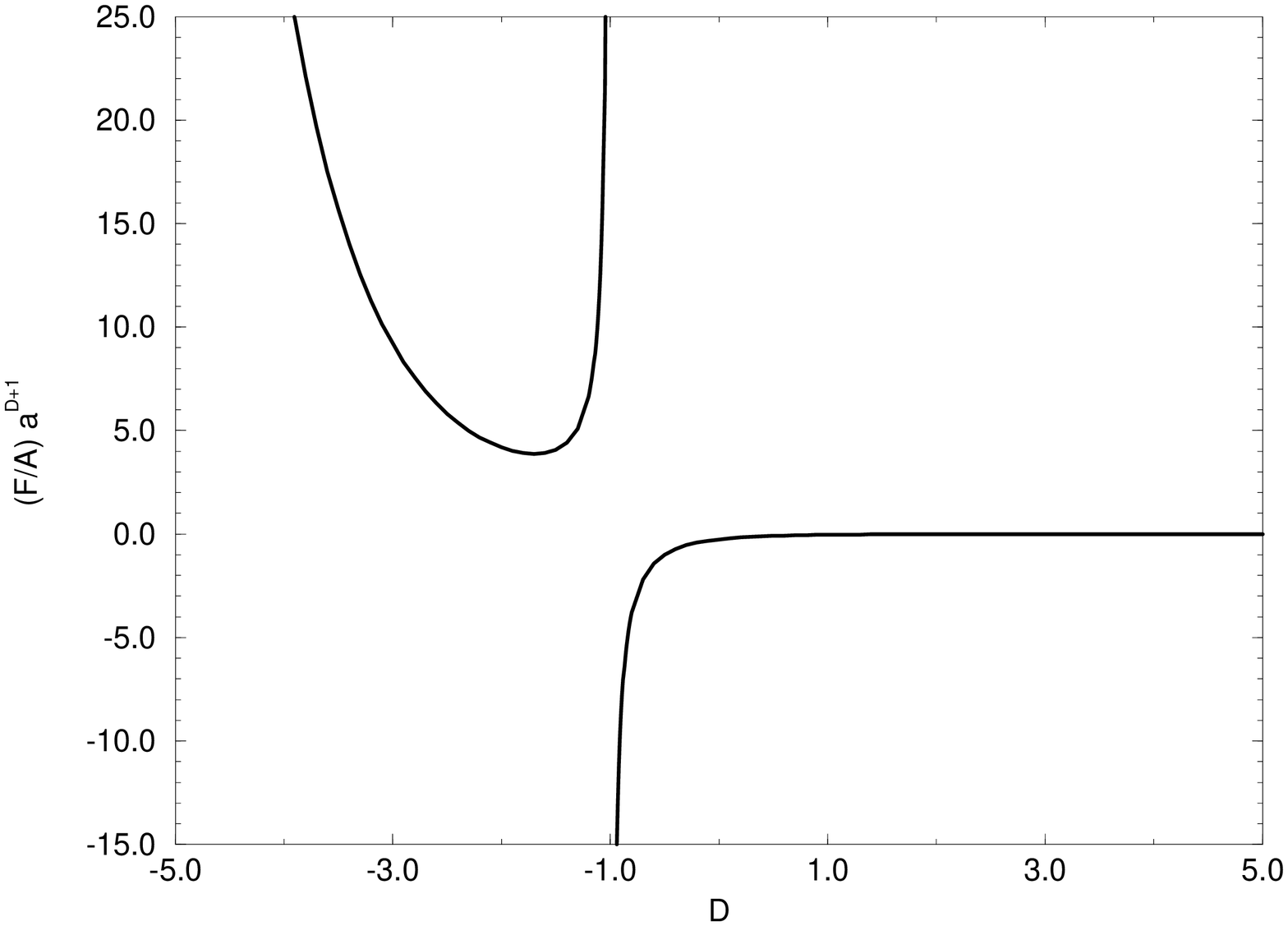,height=5in,width=6.5in,angle=270}}
\caption{A plot of the Casimir force per unit area $F/A$ in
(\protect\ref{0})
for $-5<D<5$ for the case of a slab geometry (two parallel plates).}
\label{fig1}
\end{figure}

\begin{figure}
\centerline{\psfig{figure=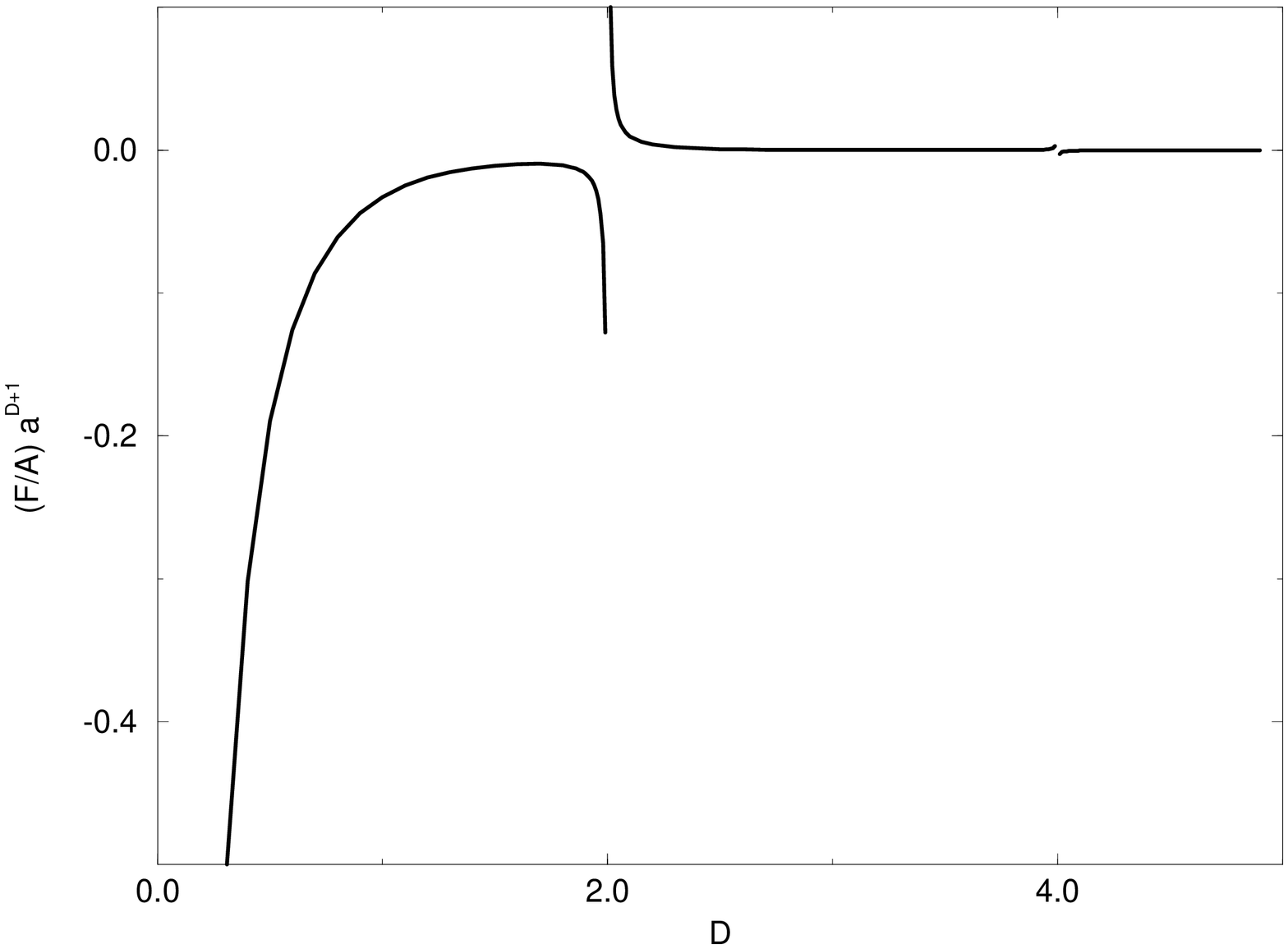,height=5in,width=6.5in,angle=270}}
\caption{A plot of the Casimir force per unit area $F/A$
for $0<D<5$ on a spherical shell.}
\label{fig2}
\end{figure}

\begin{figure}
\centerline{\psfig{figure=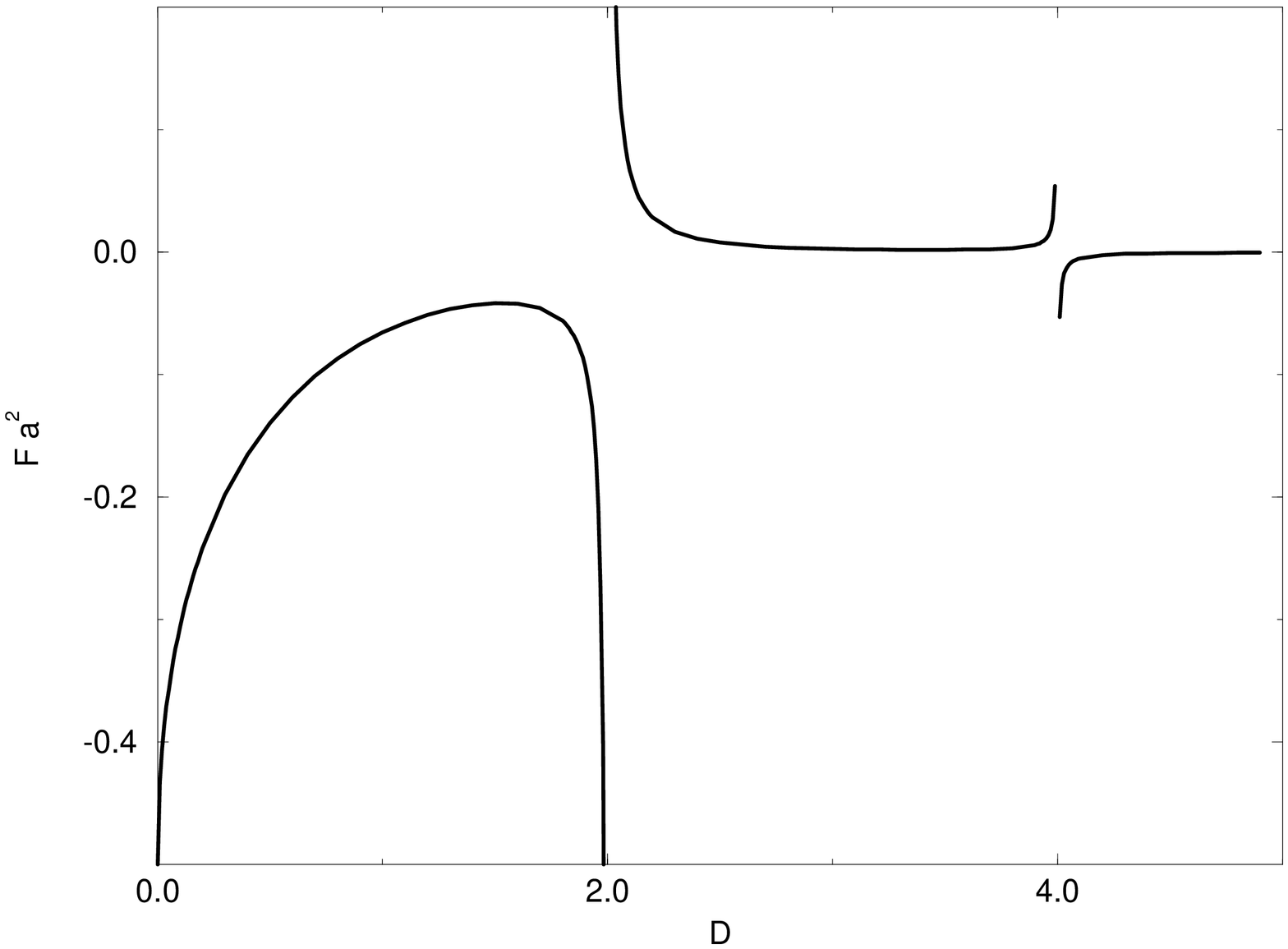,height=5in,width=6.5in,angle=270}}
\caption{A plot of the Casimir force $F$ for $0<D<5$ on a spherical
shell.}
\label{fig3}
\end{figure}

\begin{figure}
\centerline{\psfig{figure=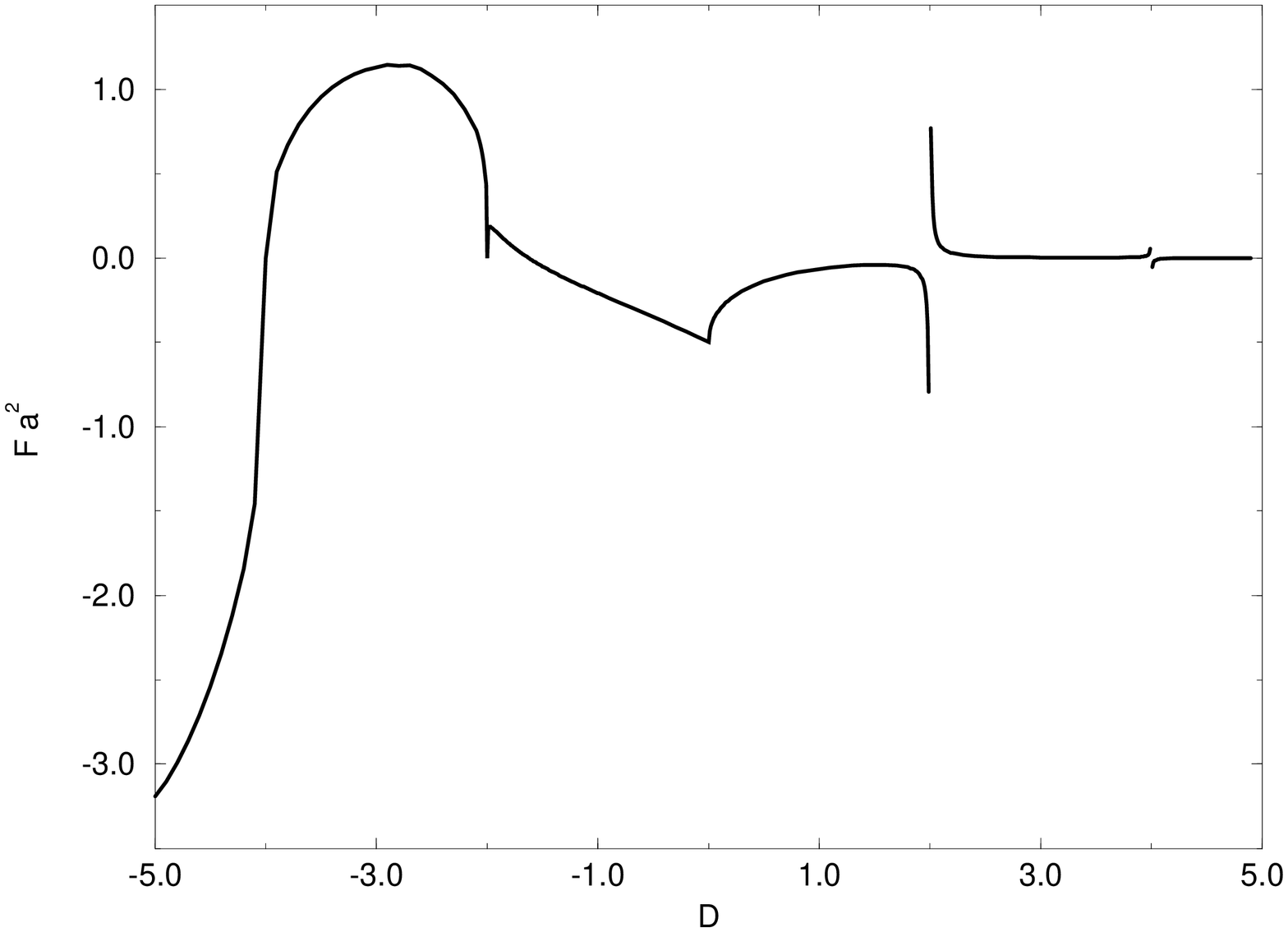,height=5in,width=6.5in,angle=270}}
\caption{A plot of the Casimir force $F$ for $-5<D<5$ on a spherical
shell.
For $D<0$ the force $F$ is complex and we have plotted ${\rm Re}\,
F$.}
\label{fig4}
\end{figure}


\begin{references}
\bibitem{BBL1}
C. M. Bender, S. Boettcher, and L. Lipatov, Phys.~Rev.~Lett. {\bf
168}, 3674
(1992).

\bibitem{BBL2}
C. M. Bender, S. Boettcher, and L. Lipatov, Phys.~Rev.~D {\bf 146},
5557 (1992).

\bibitem{Fermitalk}
C. M. Bender, in {\sl The Fermilab Meeting DPF 92,} (World
Scientific,
Singapore, 1993), Vol.~2, pp.~1549--1551.

\bibitem{finf}
C. M. Bender and S. Boettcher, J. Math.~Phys.~{\bf 135}, 1914 (1994).

\bibitem{Ising1}
C. M. Bender and S. Boettcher, Phys.~Rev.~D {\bf 148}, 4919 (1993).

\bibitem{Ising2}
C. M. Bender and S. Boettcher, submitted to Journal of Mathematical
Physics.

\bibitem{dinv}
C. M. Bender, L. D. Mlodinow, and N. Papanicolaou, Phys.~Rev.~A {\bf
25}, 1305
(1982); D. Z. Goodson and D. K. Watson, Phys.~Rev.~A {\bf 48}, 2668
(1993); D.
R. Herschbach, in {\sl Dimensional Scaling in Chemical Physics},
ed.~by D. R.
Herschbach, J. Avery, and O. Goscinski (Kluwer, Dordrecht, 1993),
pp.~7--59.

\bibitem{walk1}
C. M. Bender, S. Boettcher, and L. R. Mead, J. Math.~Phys.~{\bf 135},
368
(1994).

\bibitem{walk2}
C. M. Bender, S. Boettcher, and M. Moshe, Journal of Mathematical
Physics (in
press).

\bibitem{wolf}
J. Ambj{\o}rn and S. Wolfram, Ann.~Phys.~(N.Y.) {\bf 147}, 1 (1983).

\bibitem{mil1}
K. A. Milton, L. L. DeRaad, Jr., and J. Schwinger, Ann.~Phys.~(N.Y.)
{\bf 115},
388 (1978).

\bibitem{mil2}
L. L. DeRaad, Jr. and K. A. Milton, Ann.~Phys.~(N.Y.) {\bf 136}, 229
(1981).

\bibitem{mil3}
K. A. Milton and Y. J. Ng, Phys.~Rev.~D {\bf 46}, 842 (1992).

\bibitem{elcid}
One might wonder whether using an alternative stress-energy tensor
would give
a different Casimir force. For example, one might use the conformally
invariant
``new improved'' stress-energy tensor of C. G. Callan, Jr., S.
Coleman, and
R. Jackiw, Ann.~Phys.~{\bf 59}, 42 (1970). However, a direct
calculation reveals
that the magnitude of the Casimir force is {\sl independent\/} of the
choice of
stress-energy tensor.

\bibitem{NBS}
{\sl Handbook of Mathematical Functions}, ed. by M. Abramowitz and I.
A. Stegun
(National Bureau of Standards, Washington, D.C., 1964), Chap.~22.

\bibitem{NBS2}
{\sl Ibid.}, Chap.~9.

\bibitem{tim}
T. H. Boyer, Phys.~Rev.~{\bf 174}, 1764 (1968). See also R. Balian
and R.
Duplantier, Ann.~Phys.~(N.Y.) {\bf 112}, 165 (1978).

\bibitem{kim}
K. A. Milton, Ann.~Phys.~(N.Y.) {\bf 150}, 432 (1983).
\end{references}
\end{document}